\documentclass{article}
\usepackage{amsmath}
\usepackage{graphicx}
\begin{document}
\title{
Ward Identities for Cooper Pairs
\footnote{
APPENDIX has been added in v2.}
}

\author{
O. Narikiyo
\footnote{
Department of Physics, 
Kyushu University, 
Fukuoka 812-8581, 
Japan}
\footnote{
narikiyo@phys.kyushu-u.ac.jp}
}

\date{
(Aug. 17, 2011)
}

\maketitle
\begin{abstract}
Ward identities for Cooper pairs are derived. 
These give consistent description of 
electronic curent vertex and thermal current vertex. 
\end{abstract}
\vskip 20pt

The quest for the correct expression of 
the thermal current vertex for Cooper pairs 
has a long history but the present status is 
still controversial. 

\vskip 15pt

A phenomenological expression 
on the basis of Ginzburg-Landau (GL) theory~\cite{USH,LV} 
is naively expected to be reliable. 
The GL theory relates the electronic current vertex ${\vec J}^e$ 
and the thermal current vertex ${\vec J}^Q$ for Cooper pairs as 
\begin{equation}
{\vec J}^Q = {1 \over 2e}
\left( {i\omega_m + {i\omega_\nu \over 2}} \right) {\vec J}^e, 
\end{equation}
in the limits of long wavelength and low frequency 
where incoming and outgoing Cooper pairs have bosonic thermal frequencies 
$i\omega_m$ and $i\omega_m + i\omega_\nu$ and 
$e$ is the charge of an electron ($e<0$). 

\vskip 15pt

On the other hand, 
the most recent works~\cite{SSVG,LNV} in this field 
relate these two vertices as 
\begin{equation}
{\vec J}^Q = {1 \over e}
\left( {i\omega_m + {i\omega_\nu \over 2}} \right) {\vec J}^e. 
\end{equation}
However the origin of the factor 2 has not been explained convincingly. 

\vskip 15pt

In this Short Note we try to obtain the correct relation 
between ${\vec J}^e$ and ${\vec J}^Q$ 
on the basis of Ward identities. 

\vskip 15pt

First 
we review the derivation~\cite{Sch} of the Ward identity 
for electronic current vertex. 
The vertex function is defined as 
\begin{equation}
\Lambda_\mu^e(x,y,z) = \langle T_\tau 
\{ j_\mu^e(z) \psi_\uparrow(x) \psi_\uparrow^\dag(y) \} \rangle,
\end{equation}
and under the charge-current conservation 
($\sum_{\mu=0}^3 {\partial \over \partial z_\mu} j_\mu^e(z) = 0$)
its divergence is transformed into 
\begin{eqnarray}
- i 
\sum_{\mu=0}^3 {\partial \over \partial z_\mu} \Lambda_\mu^e(x,y,z) &=& 
\langle T_\tau 
\{ [j_0^e(z), \psi_\uparrow(x)] \psi_\uparrow^\dag(y) \} \rangle 
\delta(\tau_z - \tau_x) \nonumber \\ 
&+& 
\langle T_\tau 
\{ \psi_\uparrow(x) [j_0^e(z), \psi_\uparrow^\dag(y)] \} \rangle 
\delta(\tau_z - \tau_y), 
\end{eqnarray}
where 
$\langle \cdot\cdot\cdot \rangle$ represents the thermal average, 
$T_\tau$ is the time-ordering operator 
with respect to the imaginary time $ \tau_z = i z_0 $ 
and $z=({\vec z},z_0)$ 
with the coordinate vector ${\vec z}=(z_1,z_2,z_3)$ 
and the real time $z_0$. 
Here $\psi_\uparrow(x)$ and $\psi_\uparrow^\dag(y)$ are 
annihilation and creation operators of $\uparrow$-spin electron. 
The zeroth component of the electron current $j_0^e(z)$ is given by 
$j_0^e(z) = 
e \psi_\uparrow^\dag(z) \psi_\uparrow(z) + 
e \psi_\downarrow^\dag(z) \psi_\downarrow(z)$. 
Using the commutation relation 
$[j_0^e(z), \psi_\uparrow^\dag(z)] = e \psi_\uparrow^\dag(z)$ and 
$[j_0^e(z), \psi_\uparrow(z)] = - e \psi_\uparrow(z)$ and 
introducing the Fourier transform, we obtain the Ward identity 
\begin{equation}
\sum_{\mu=0}^3 k_\mu \Gamma_\mu^e(p,k) = 
e G^{-1}(p) - e G^{-1}(p+k), 
\end{equation}
for the electronic current vertex 
where 
$\Lambda_\mu^e(p,k) = G(p) \Gamma_\mu^e(p,k) G(p+k)$ 
and $G(p)$ is the electron propagator 
with four-momentum $p=({\vec p},p_0)$. 
The zeroth components $p_0$ and $k_0$ are 
fermionic ($p_0=-i\varepsilon_n$) and 
bosonic ($k_0=-i\omega_\nu$) frequencies. 

The extension of this Ward identity to the case of Cooper pairs 
is straightforward. 
In the following 
we consider the Cooper pair of s-wave pairing 
in the case of local attractive interaction. 
Since we discuss the normal metallic phase ($T>T_c$), 
the propagator of Cooper pairs is a fluctuation propagator. 
Replacing 
$\psi_\uparrow(x)$ by $\psi_\downarrow(x) \psi_\uparrow(x)$ and 
$\psi_\uparrow^\dag(y)$ by $\psi_\uparrow^\dag(y) \psi_\downarrow^\dag(y)$ 
and using the commutation relation 
$[j_0^e(z), \psi_\uparrow^\dag(z) \psi_\downarrow^\dag(z)] = 
2e \psi_\uparrow^\dag(z) \psi_\downarrow^\dag(z)$ and 
$[j_0^e(z), \psi_\downarrow(z) \psi_\uparrow(z)] = 
-2e \psi_\downarrow(z) \psi_\uparrow(z)$, 
we obtain the Ward identity 
\begin{equation}
\sum_{\mu=0}^3 k_\mu \Delta_\mu^e(q,k) = 
2e D^{-1}(q) - 2e D^{-1}(q+k), 
\end{equation}
for Cooper pairs 
where 
$\Delta_\mu^e$ is the counterpart of $\Gamma_\mu^e$ and 
$D(q)$ is the Cooper-pair fluctuation propagator 
with four-momentum $q=({\vec q},q_0)$ 
whose zeroth component $q_0$ is a bosonic frequency ($q_0=-i\omega_m$). 
It should be noted that the factor $2e$ represents 
the charge carried by a Cooper pair and 
is automatically taken into account by the commutation relations. 
This point was missed in an early guess~\cite{Tsu} 
of the Ward identity for Cooper pairs. 

\vskip 15pt

Second 
we review the derivation~\cite{Ono} of the Ward identity 
for thermal current vertex. 
Since the essence of the derivation is the same 
as the electronic current vertex, 
this review is short. 
In the limit of vanishing external momentum, ${\vec k}\rightarrow 0$, 
the Ward identity is obtained as 
\begin{equation}
\sum_{\mu=0}^3 k_\mu \Gamma_\mu^Q(p+k,p) = 
p_0 G^{-1}(p+k) - (p_0+k_0) G^{-1}(p), 
\end{equation}
for electrons. 
Here $p_0$ and $p_0+k_0$ result from 
the Fourier transform of the time-derivative of 
annihilation and creation operators of $\uparrow$-spin electron. 
The time-derivative results from 
the commutation relation between the zeroth component of the thermal current 
and annihilation or creation operator, 
since the zeroth component for ${\vec k}\rightarrow 0$ 
is the Hamiltonian of the system. 

The extension of this Ward identity to the case of Cooper pairs 
is also straightforward. 
The Ward identity for thermal current vertex for Cooper pairs is 
\begin{equation}
\sum_{\mu=0}^3 k_\mu \Delta_\mu^Q(q+k,q) = 
q_0 D^{-1}(q+k) - (q_0+k_0) D^{-1}(q). 
\end{equation}

\vskip 15pt

Finally 
our Ward identities eqs. (6) and (8) are consistent with 
the GL result eq. (1). 

\vskip 15pt

This work arose from discussions with 
Kazumasa Miyake, Yukinobu Fujimoto and Shinji Watanabe 
at Osaka University. 

\vskip 30pt

\noindent
{\Large {\bf APPENDIX}}

\vskip 10pt

\noindent
In this APPENDIX the derivation of the Ward identity for Cooper pairs 
is explained in detail. 
\vskip 20pt

\section{Introduction}

The Ward identity for electric current vertex is explained 
in Schrieffer's textbook~\cite{Sch}. 

The Ward identity for heat current vertex is derived 
in Ono's paper~\cite{Ono}. 

If we understand these discussions well, 
we can reach the Ward identity for Cooper pairs with little effort. 
Thus I review these works first and 
cast the results into those for Cooper pairs.

\section{Ward Identity for Electric Current Vertex}

First 
we review the derivation~\cite{Sch} of the Ward identity 
for electric current vertex. 
We mainly discuss the case of zero temperature and 
cast the zero-temperature result into that of finite temperature. 

We consider the three-point function $\Lambda_\mu^e$ 
($\mu = 1,2,3,0$) defined as 
\begin{equation}
\Lambda_\mu^e(x,y,z) = \langle T 
\{ j_\mu^e(z) \psi_\uparrow(x) \psi_\uparrow^\dag(y) \} \rangle,
\end{equation}
where 
$\langle \cdot\cdot\cdot \rangle$ 
represents the expectation value in the ground state, 
$T$ is the time-ordering operator and $z=({\vec z},z_0)$ 
with coordinate vector ${\vec z}=(z_1,z_2,z_3)$ and time $z_0$. 
Here $\psi_\uparrow(x)$ and $\psi_\uparrow^\dag(y)$ are 
annihilation and creation operators of $\uparrow$-spin electron. 
The electric current $j_\mu^e$ 
obeys the charge-conservation law 
\begin{equation}
\sum_{\mu=0}^3 {\partial \over \partial z_\mu} j_\mu^e(z) = 0. \label{cc-law}
\end{equation}
Especially the electric charge $j_0^e$ is given by 
\begin{equation}
j_0^e(z) = 
e \psi_\uparrow^\dag(z) \psi_\uparrow(z) + 
e \psi_\downarrow^\dag(z) \psi_\downarrow(z), 
\end{equation}
where $e$ is the charge of an electron ($e<0$). 
The time ordering of three operators results in 
the summation of $3!$ terms as 
\begin{align}
\Lambda_\mu^e(x,y,z) &= 
\langle j_\mu^e(z) \psi_\uparrow(x) \psi_\uparrow^\dag(y) \rangle 
\theta(z_0-x_0)\theta(x_0-y_0) \nonumber \\ 
&- 
\langle j_\mu^e(z) \psi_\uparrow^\dag(y) \psi_\uparrow(x) \rangle 
\theta(z_0-y_0)\theta(y_0-x_0) \nonumber \\ 
&+ 
\langle \psi_\uparrow(x) j_\mu^e(z) \psi_\uparrow^\dag(y) \rangle 
\theta(x_0-z_0)\theta(z_0-y_0) \nonumber \\ 
&- 
\langle \psi_\uparrow^\dag(y) j_\mu^e(z) \psi_\uparrow(x) \rangle 
\theta(y_0-z_0)\theta(z_0-x_0) \nonumber \\ 
&+ 
\langle \psi_\uparrow(x) \psi_\uparrow^\dag(y) j_\mu^e(z) \rangle 
\theta(x_0-y_0)\theta(y_0-z_0) \nonumber \\ 
&- 
\langle \psi_\uparrow^\dag(y) \psi_\uparrow(x) j_\mu^e(z) \rangle 
\theta(y_0-x_0)\theta(x_0-z_0), 
\label{3!} 
\end{align}
where 
$\theta(x)$ is the unit step function. 
Thus the time-derivative of $\Lambda_\mu^e$ results in 
\begin{align}
{\partial \over \partial z_0} \Lambda_0^e(x,y,z) = 
\delta(z_0-x_0) \Bigl( 
\theta(x_0-y_0) 
&\langle [ j_0^e(z), \psi_\uparrow(x) ] \psi_\uparrow^\dag(y) \rangle 
\nonumber \\ 
- \theta(y_0-x_0) 
&\langle \psi_\uparrow^\dag(y) [ j_0^e(z), \psi_\uparrow(x) ] \rangle 
\Bigr) \nonumber \\ 
+ 
\delta(z_0-y_0) \Bigl( 
\theta(x_0-y_0) 
&\langle \psi_\uparrow(x) [ j_0^e(z), \psi_\uparrow^\dag(y) ] \rangle 
\nonumber \\ 
- \theta(y_0-x_0) 
&\langle [ j_0^e(z), \psi_\uparrow^\dag(y) ] \psi_\uparrow(x) \rangle 
\Bigr) \nonumber \\ 
+ 
&\langle T 
\left\{ {\partial j_0^e(z) \over \partial z_0} 
\psi_\uparrow(x) \psi_\uparrow^\dag(y) \right\} \rangle . 
\label{step-delta} 
\end{align}
Using again the time ordering 
the divergence of $\Lambda_\mu^e$ is expressed as 
\begin{align}
\sum_{\mu=0}^3 {\partial \over \partial z_\mu} \Lambda_\mu^e(x,y,z) = 
& \langle T 
\{ [j_0^e(z), \psi_\uparrow(x)] \psi_\uparrow^\dag(y) \} \rangle 
\delta(z_0-x_0) \nonumber \\ 
+ 
& \langle T 
\{ \psi_\uparrow(x) [j_0^e(z), \psi_\uparrow^\dag(y)] \} \rangle 
\delta(z_0-y_0) \nonumber \\ 
+ 
& \langle T 
\left\{ \sum_{\mu=0}^3 {\partial j_\mu^e(z) \over \partial z_\mu} 
\psi_\uparrow(x) \psi_\uparrow^\dag(y) \right\} \rangle. 
\label{divLambda} 
\end{align}
The last term on the right-hand side vanishes 
due to the charge-conservation law, eq.(\ref{cc-law}). 
Only equal space-time commutation relations are non-vanishing, 
\begin{equation}
[j_0^e(z), \psi_\uparrow^\dag(z)] = e \psi_\uparrow^\dag(z), 
\end{equation}
and 
\begin{equation}
[j_0^e(z), \psi_\uparrow(z)] = - e \psi_\uparrow(z), 
\end{equation}
so that the non-vanishing contribution becomes 
\begin{align}
\sum_{\mu=0}^3 {\partial \over \partial z_\mu} \Lambda_\mu^e(x,y,z) = 
& - e \langle T \{ \psi_\uparrow(x) \psi_\uparrow^\dag(y) \} \rangle 
\delta^4(z-x) 
\nonumber \\ 
& + e \langle T \{ \psi_\uparrow(x) \psi_\uparrow^\dag(y) \} \rangle 
\delta^4(z-y). 
\end{align}
Introducing the electron propagator $G(x,y)$ as 
\begin{equation}
G(x,y) = 
- i \langle T \{ \psi_\uparrow(x) \psi_\uparrow^\dag(y) \} \rangle, 
\label{G(t)}
\end{equation}
this relation is written into 
\begin{equation}
\sum_{\mu=0}^3 {\partial \over \partial z_\mu} \Lambda_\mu^e(x,y,z) = 
- i e G(x,y) \delta^4(z-x) + i e G(x,y) \delta^4(z-y), \label{LG}
\end{equation}
Assuming the translational invariance 
we set $y=0$ and introduce the Fourier transform as 
\begin{equation}
\Lambda_\mu^e(p,k) = \int d^4 x e^{-ipx} \int d^4 z e^{-ikz} 
\langle T 
\{ j_\mu^e(z) \psi_\uparrow(x) \psi_\uparrow^\dag(0) \} \rangle, 
\end{equation}
where the four-momentum is defined as 
$p=({\vec p},p_0)$ and $k=({\vec k},k_0)$. 
The left-hand side of eq.~(\ref{LG}) is evaluated as 
\begin{equation}
\sum_{\mu=0}^3 {\partial \over \partial z_\mu} \Lambda_\mu^e(x,0,z) = 
\int {d^4 p \over (2\pi)^4} e^{ipx} \int {d^4 k \over (2\pi)^4}e^{ikz} 
\sum_{\mu=0}^3 i k_\mu \Lambda_\mu^e(p,k), 
\end{equation}
and the right-hand side is transformed as 
\begin{align}
\int d^4 x e^{-ipx} \int d^4 z e^{-ikz} 
\Bigl(
- G(x,0) \delta^4(z-x) + G(x,0) \delta^4(z) 
\Bigr) \nonumber \\ 
= - G(p+k) + G(p), 
\end{align}
where 
\begin{equation}
G(p) = 
\int d^4 x e^{-ipx} G(x,0). 
\end{equation}
Therefore we obtain 
\begin{equation}
\sum_{\mu=0}^3 k_\mu \Lambda_\mu^e(p,k)
= e G(p) - e G(p+k). 
\end{equation}
The vertex function $\Gamma_\mu^e$ is introduced as 
\begin{equation}
\Lambda_\mu^e(p,k) = i G(p) \cdot \Gamma_\mu^e(p,k) \cdot i G(p+k), 
\label{LiG}
\end{equation}
in accordance with the definiton of the Green function, eq.~(\ref{G(t)}). 
Then the Ward identity for the electric current vertex 
is given by 
\begin{equation}
\sum_{\mu=0}^3 k_\mu \Gamma_\mu^e(p,k) = 
e G^{-1}(p) - e G^{-1}(p+k). \label{WI-e} 
\end{equation}
Since the Fourier transform is introduced as 
\begin{equation}
px = p_1x_1 + p_2x_2 + p_3x_3 - \epsilon t, 
\end{equation}
where $\epsilon$ is the energy and $t$ is the time, 
$x_0 = t$ and $p_0 = - \epsilon$ 
(and in the same manner $k_0 = - \omega$ 
with $\omega$ being the energy of the external field) 
in the zero-temperatute formalism. 

Here we check the limiting case of eq.~(\ref{WI-e}). 
If we replace the full Green function $G(p)$ 
by the free Green function $G_0(p)$ and set $k_0=0$, we obtain 
\begin{equation}
\sum_{\mu=1}^3 k_\mu \Gamma_\mu^e(p,k) = 
{ e \over m } {\vec k} \cdot ( {\vec p}+{{\vec k} \over 2} ), 
\end{equation}
where the free dispersion $\epsilon_{\vec p}={\vec p}^2 / 2m$ 
is employed with $m$ being the mass of electron. 
This relation means 
\begin{equation}
\Gamma_\mu^e(p,0) = e v_\mu, 
\end{equation}
where the right-hand side is the proper electric current vertex 
with the electron velocity ${\vec v} = {\vec p}/m $. 

In the finite-temperature formalism 
we employ the time-ordering operator $T_\tau$ and 
consider the three-point function
\begin{equation}
\Lambda_\mu^e(x,y,z) = \langle T_\tau 
\{ j_\mu^e(z) \psi_\uparrow(x) \psi_\uparrow^\dag(y) \} \rangle, 
\end{equation}
where the real time $z_0$ and the imaginary time $\tau_z$ 
is related by $\tau_z = i z_0$ and 
$\langle \cdot\cdot\cdot \rangle$ represents the thermal average. 
Taking the charge-conservation law, eq.(\ref{cc-law}), 
into account we obtain 
\begin{align}
- i \sum_{\mu=0}^3 {\partial \over \partial z_\mu} \Lambda_\mu^e(x,y,z) = 
& \langle T_\tau 
\{ [j_0^e(z), \psi_\uparrow(x)] \psi_\uparrow^\dag(y) \} \rangle 
\delta(\tau_z - \tau_x) \nonumber \\ 
+ 
& \langle T_\tau 
\{ \psi_\uparrow(x) [j_0^e(z), \psi_\uparrow^\dag(y)] \} \rangle 
\delta(\tau_z - \tau_y), 
\end{align}
instead of eq.~(\ref{divLambda}). 
Here we have used the relation 
\begin{equation}
{\partial \over \partial \tau_z} = 
- i {\partial \over \partial z_0}. \label{del-tau}
\end{equation}
The finite-temperature vertex function $\Gamma_\mu^e$ is introduced as 
\begin{equation}
\Lambda_\mu^e(p,k) = [-G(p)] \cdot \Gamma_\mu^e(p,k) \cdot [-G(p+k)], 
\end{equation}
in accordance with the definiton of the thermal Green function 
\begin{equation}
G(x,y) = 
- \langle T_\tau \{ \psi_\uparrow(x) \psi_\uparrow^\dag(y) \} \rangle. 
\label{G(tau)}
\end{equation}
Then the resulting Ward identity is the same form as eq.~(\ref{WI-e}) 
in the case of zero temperature. 
For the finite-temperature Ward identity 
the zeroth component of the four-momentum is 
$p_0 = - i \varepsilon_n$ with fermionic thermal frequency $\varepsilon_n$
and $k_0 = - i \omega_\nu$ with bosonic thermal frequency $\omega_\nu$. 

The extension of this Ward identity to the case of Cooper pairs 
is straightforward. 
In the following 
we consider the Cooper pair of s-wave pairing 
in the case of local attractive interaction. 
Replacing 
$\psi_\uparrow(x)$ by 
$\Psi(x) = \psi_\downarrow(x) \psi_\uparrow(x)$ and 
$\psi_\uparrow^\dag(y)$ by 
$\Psi^\dag(y) = \psi_\uparrow^\dag(y) \psi_\downarrow^\dag(y)$ 
we consider the three-point function $M_\mu^e$ as 
\begin{equation}
M_\mu^e(x,y,z) = \langle T 
\{ j_\mu^e(z) \Psi(x) \Psi^\dag(y) \} \rangle,
\end{equation}
where 
$\Psi(x)$ and $\Psi^\dag(y)$ are 
annihilation and creation operators of a Cooper pair 
which has a bosonic character. 
Using the commutation relation 
\begin{equation}
[j_0^e(z), \Psi^\dag(z)] = 2e \Psi^\dag(z), 
\end{equation}
and 
\begin{equation}
[j_0^e(z), \Psi(z)] = -2e \Psi(z), 
\end{equation}
the divergence of $M_\mu^e$ is expressed as 
\begin{align}
\sum_{\mu=0}^3 {\partial \over \partial z_\mu} M_\mu^e(x,y,z) = 
& - 2e \langle T \{ \Psi(x) \Psi^\dag(y) \} \rangle \delta^4(z-x) 
\nonumber \\ 
& + 2e \langle T \{ \Psi(x) \Psi^\dag(y) \} \rangle \delta^4(z-y), 
\end{align}
by repeating the same calculations as 
eqs.~(\ref{3!}), (\ref{step-delta}) and (\ref{divLambda}). 
Here the difference between fermion and boson is handled 
solely by the time-ordering operator $T$ so that 
the expression of the divergence is common to fermion and boson. 
Introducing the Cooper-pair propagator $D(x,y)$ as 
\begin{equation}
D(x,y) = 
- i \langle T \{ \Psi(x) \Psi^\dag(y) \} \rangle, 
\end{equation}
we obtain the Ward identity 
\begin{equation}
\sum_{\mu=0}^3 k_\mu \Delta_\mu^e(q,k) = 
2e D^{-1}(q) - 2e D^{-1}(q+k), \label{WIC-e}
\end{equation}
for Cooper pairs 
where 
$\Delta_\mu^e$ is the counterpart of $\Gamma_\mu^e$ and 
$D(q)$ is the Fourier transform of $D(x,0)$ 
with four-momentum $q=({\vec q},q_0)$. 

Although the above derivation for Cooper pairs 
is formulated at zero temperature, 
eq.~(\ref{WIC-e}) also holds at finite temperature 
with $q_0$ being a bosonic thermal frequency ($q_0 = - i \omega_m$). 
We are mainly interested in the normal metallic phase ($T>T_c$), 
the Cooper-pair propagator is a fluctuation propagator in this case. 

It should be noted that the factor $2e$ represents 
the charge carried by a Cooper pair and 
is automatically taken into account by the commutation relation. 

\section{Ward Identity for Heat Current Vertex}

First 
we review the derivation~\cite{Ono} of the Ward identity 
for heat current vertex. 
We consider the three-point function $\Lambda_\mu^Q$ 
defined as 
\begin{equation}
\Lambda_\mu^Q(x,y,z) = \langle T 
\{ j_\mu^Q(z) \psi_\uparrow(x) \psi_\uparrow^\dag(y) \} \rangle,
\end{equation}
where $j_\mu^Q$ is the heat current. 
The heat current $j_\mu^Q$ 
obeys the energy-conservation law 
\begin{equation}
\sum_{\mu=0}^3 {\partial \over \partial z_\mu} j_\mu^Q(z) = 0. \label{ec-law}
\end{equation}
Assuming the translational invariance 
we set 
\begin{equation}
\Lambda_\mu^Q(x,y,z) = \Lambda_\mu^Q(x-y,z-x), 
\end{equation}
so that the Fourier transform becomes 
\begin{equation}
\int d^4 z e^{-ikz} \int d^4 x e^{-ip'x} \int d^4 y e^{ipy} 
\Lambda_\mu^Q(x,y,z) 
= \Lambda_\mu^Q(p,p-k) 
(2\pi)^4 \delta^4(-k-p'+p), 
\end{equation}
where 
\begin{equation}
\Lambda_\mu^Q(p,p-k) = 
\int d^4 (x-y) e^{-ip(x-y)} \int d^4 (z-x) e^{-ik(z-x)} 
\Lambda_\mu^Q(x-y,z-x), \label{trans-inv-int} 
\end{equation}
and 
$\delta(-k-p'+p)$ 
represents the conservation of four-momentum. 
The divergence of $\Lambda_\mu^Q$ becomes 
\begin{align}
\sum_{\mu=0}^3 {\partial \over \partial z_\mu} \Lambda_\mu^Q(x,y,z) = 
& \langle T 
\{ [j_0^Q(z), \psi_\uparrow(x)] \psi_\uparrow^\dag(y) \} \rangle 
\delta(z_0-x_0) \nonumber \\ 
+ 
& \langle T 
\{ \psi_\uparrow(x) [j_0^Q(z), \psi_\uparrow^\dag(y)] \} \rangle 
\delta(z_0-y_0), \label{divLamQxyz} 
\end{align}
under the energy-conservation law, eq.(\ref{ec-law}). 
The left-hand side is evaluated as 
\begin{equation}
\sum_{\mu=0}^3 {\partial \over \partial z_\mu} \Lambda_\mu^Q(x-y,z-x) = 
\int {d^4 p \over (2\pi)^4} e^{ip(x-y)} \int {d^4 k \over (2\pi)^4}e^{ik(z-x)} 
\sum_{\mu=0}^3 i k_\mu \Lambda_\mu^Q(p,p-k), 
\end{equation}
so that the Fourier transform satisfies 
\begin{align}
\sum_{\mu=0}^3 i k_\mu \Lambda_\mu^Q(p,p-k) = 
\int d & (x_0-y_0) e^{-ip_0(x_0-y_0)} \int d (z_0-x_0) e^{-ik_0(z_0-x_0)} 
\nonumber \\ 
\times \Bigl(
& \langle T 
\{ [j_{\vec k}^Q(x_0), a_{{\vec p}-{\vec k}}(x_0)] a_{\vec p}^\dag(y_0) \} 
\rangle \delta(z_0-x_0) \nonumber \\ 
+ 
& \langle T 
\{ a_{{\vec p}-{\vec k}}(x_0) [j_{\vec k}^Q(y_0), a_{\vec p}^\dag(y_0)] \} 
\rangle \delta(z_0-y_0) \Bigr), \label{divLambdaQ} 
\end{align}
where 
\begin{equation}
j_0^Q(z) = 
\sum_{\vec k} e^{i{\vec k}\cdot{\vec z}} j_{\vec k}^Q(z_0),\ \ \ 
\psi_\uparrow(x) = 
\sum_{\vec p'} e^{i{\vec p'}\cdot{\vec x}} a_{\vec p'}(x_0),\ \ \ 
\psi_\uparrow^\dag(y) = 
\sum_{\vec p} e^{-i{\vec p}\cdot{\vec y}} a_{\vec p}^\dag(y_0), 
\end{equation}
with 
\begin{equation}
\sum_{\vec k} \equiv \int { d{\vec k} \over (2\pi)^3 }. 
\end{equation}

Then we evaluate the commutation relations. 
For such a purpose we introduce the Hamiltonian density $h(z)$ 
for an isotropic system as 
$ h(z) = h^{\rm kin}({\vec z}) + h^{\rm int}({\vec z})$ at time $z_0$ 
where 
\begin{equation}
h^{\rm kin}({\vec z}) = {1 \over 2m} \sum_{\sigma} 
\nabla \psi_\sigma^\dag({\vec z}) \cdot \nabla \psi_\sigma({\vec z}), 
\end{equation}
and
\begin{equation}
h^{\rm int}({\vec z}) = 
{ 1 \over 2 } \sum_{\sigma} \sum_{\sigma'} \int d{\vec z'} 
\psi_\sigma^\dag({\vec z}) \psi_{\sigma'}^\dag({\vec z'})
V({\vec z}-{\vec z'})
\psi_{\sigma'}({\vec z'}) \psi_\sigma({\vec z}), 
\end{equation}
with a general interaction strength $V({\vec z}-{\vec z'})$ 
of two-body interaction 
and spin $\sigma, \sigma' = \uparrow, \downarrow$. 
Within this paragraph the time of all operators is $z_0$. 
This Hamiltonian density $h({\vec z})$ is 
the zeroth component of the heat current $j_0^Q(z)$. 
The Fourier component 
\begin{equation}
j_{\vec k}^Q \equiv j_{\vec k}^Q(z_0) = 
\int d{\vec z} e^{-i{\vec k}\cdot{\vec z}} j_0^Q(z), 
\end{equation}
is decomposed as 
$ j_{\vec k}^Q = j_{\vec k}^{\rm kin} + j_{\vec k}^{\rm int}$
and 
\begin{equation}
j_{\vec k}^{\rm kin} = {1 \over 2m} \sum_{\vec p} 
({\vec p}-{\vec k})\cdot \vec p
\Bigl( a_{{\vec p}-{\vec k}}^\dag a_{\vec p} 
     + b_{{\vec p}-{\vec k}}^\dag b_{\vec p} \Bigr), 
\end{equation}
where $b_{\vec p}$ and $b_{{\vec p}-{\vec k}}^\dag$ 
are the Fourier components of
annihilation and creation operators of $\downarrow$-spin electron. 
Comparing 
\begin{equation}
[ j_{\vec k}^{\rm kin}, a_{{\vec p}-{\vec k}} ] = 
- { ({\vec p}-{\vec k})\cdot \vec p \over 2m } a_{\vec p}, \ \ \ 
[ j_{\vec k}^{\rm kin}, a_{\vec p}^\dag ] = 
{ ({\vec p}-{\vec k})\cdot \vec p \over 2m } a_{{\vec p}-{\vec k}}^\dag, 
\end{equation}
with 
\begin{equation}
[ j_{\vec k=0}^{\rm kin}, a_{\vec p} ] = 
- { {\vec p}\cdot{\vec p} \over 2m } a_{\vec p}, \ \ \ 
[ j_{\vec k=0}^{\rm kin}, a_{{\vec p}-{\vec k}}^\dag ] = 
{ ({\vec p}-{\vec k})\cdot({\vec p}-{\vec k}) \over 2m } 
a_{{\vec p}-{\vec k}}^\dag, 
\end{equation}
we can use the replacement 
\begin{equation}
[ j_{\vec k}^{\rm kin}, a_{{\vec p}-{\vec k}} ] \Rightarrow 
[ j_{\vec k=0}^{\rm kin}, a_{\vec p} ], \ \ \ 
[ j_{\vec k}^{\rm kin}, a_{\vec p}^\dag ] \Rightarrow 
[ j_{\vec k=0}^{\rm kin}, a_{{\vec p}-{\vec k}}^\dag ], 
\end{equation}
in the limit of vanishing external momentum, ${\vec k}\rightarrow 0$. 
The similar argument holds for 
$[ j_{\vec k}^{\rm int}, a_{{\vec p}-{\vec k}} ]$ and 
$[ j_{\vec k}^{\rm int}, a_{\vec p}^\dag ]$ 
so that we obtain 
\begin{equation}
[ j_{\vec k}^Q, a_{{\vec p}-{\vec k}} ] \Rightarrow 
[ H, a_{\vec p} ], \ \ \ 
[ j_{\vec k}^Q, a_{\vec p}^\dag ] \Rightarrow 
[ H, a_{{\vec p}-{\vec k}}^\dag ], 
\end{equation}
since $ j_{\vec k=0}^Q = H $ 
where 
\begin{equation}
H = \int h({\vec z})d{\vec z} , 
\end{equation}
is the Hamiltonian. 

Thus the equation of motion 
\begin{equation}
[ H, a_{\vec p}(x_0) ] = - i 
{\partial \over \partial x_0} a_{\vec p}(x_0), \ \ \ 
[ H, a_{{\vec p}-{\vec k}}^\dag(y_0) ] = - i 
{\partial \over \partial y_0} a_{{\vec p}-{\vec k}}^\dag(y_0), 
\end{equation}
can be applied to eq.~(\ref{divLambdaQ}) and the result is 
\begin{align}
\sum_{\mu=0}^3 & i k_\mu \Lambda_\mu^Q(p,p-k) = 
\int d (x_0-y_0) e^{-ip_0(x_0-y_0)} \int d (z_0-x_0) e^{-ik_0(z_0-x_0)} 
\nonumber \\ 
& \times \Bigl(
{\partial \over \partial x_0} G_{\vec p}(x_0-y_0) \delta(z_0-x_0) 
+ 
{\partial \over \partial y_0} G_{{\vec p}-{\vec k}}(x_0-y_0) \delta(z_0-y_0) 
\Bigr), \label{intGp}
\end{align}
where 
\begin{equation}
G_{\vec p}(x_0-y_0) = 
- i \langle T \{ a_{\vec p}(x_0) a_{\vec p}^\dag(y_0) \} \rangle. 
\end{equation}
Employing 
\begin{equation}
{\partial \over \partial x_0} G_{\vec p}(x_0-y_0) = 
\int { d p'_0 \over 2\pi } e^{ip'_0(x_0-y_0)} i p'_0 G_{\vec p}(p'_0), 
\label{int-ip}
\end{equation}
the first term on the right-hand side of eq.~(\ref{intGp}) becomes 
$ i p_0 G(p) $ where $ G(p) = G_{\vec p}(p_0)$. 
The second term is integrated in the same manner and becomes 
$ - i (p_0-k_0) G(p-k) $. 
Then shifting the four-momentum we obtain 
\begin{equation}
\sum_{\mu=0}^3 k_\mu \Lambda_\mu^Q(p+k,p) = 
(p_0+k_0) G(p+k) - p_0 G(p). \label{WI-shift}
\end{equation}
This relation is converted into the Ward identity 
\begin{equation}
\sum_{\mu=0}^3 k_\mu \Gamma_\mu^Q(p+k,p) = 
p_0 G^{-1}(p+k) - (p_0+k_0) G^{-1}(p), \label{WI-Q}
\end{equation}
for heat current vertex. 
Here we have used the same relation 
$ \Lambda_\mu^Q(p+k,p) = i G(p+k) \cdot \Gamma_\mu^Q(p+k,p) \cdot i G(p) $ 
as eq.~(\ref{LiG}). 

In the case of finite temperature 
the same relation as eq.~(\ref{WI-Q}) holds. 

If we only consider the local interaction, 
the derivation becomes very simple~\cite{Kon} as follows. 
Since we are mainly interested in BCS-type local interaction, 
such a derivation is sufficient for our purpose. 
Taking into account the fact, 
$ j_0^Q(z) = h(z) $, 
eq.~(\ref{divLamQxyz}) is equivalent to 
\begin{align}
\sum_{\mu=0}^3 {\partial \over \partial z_\mu} \Lambda_\mu^Q(x,y,z) = 
& \langle T 
\{ [h(x), \psi_\uparrow(x)] \psi_\uparrow^\dag(y) \} \rangle 
\delta^4(z-x) \nonumber \\ 
+ 
& \langle T 
\{ \psi_\uparrow(x) [h(y), \psi_\uparrow^\dag(y)] \} \rangle 
\delta^4(z-y). \label{divLamQ-h} 
\end{align}
If the interaction among electrons is local, then 
$ [h(x), \psi_\uparrow(x)] = [H, \psi_\uparrow(x)] $ and 
$ [h(y), \psi_\uparrow^\dag(y)] = [H, \psi_\uparrow^\dag(y)] $. 
Thus using 
\begin{equation}
[h(x), \psi_\uparrow(x)] = - i 
{\partial \over \partial x_0} \psi_\uparrow(x), \ \ \ 
[h(y), \psi_\uparrow^\dag(y)] = - i 
{\partial \over \partial y_0} \psi_\uparrow^\dag(y), 
\end{equation}
eq.~(\ref{divLamQ-h}) becomes 
\begin{equation}
\sum_{\mu=0}^3 {\partial \over \partial z_\mu} \Lambda_\mu^Q(x,y,z) = 
{\partial \over \partial x_0} G(x,y) \delta^4(z-x) 
+ 
{\partial \over \partial y_0} G(x,y) \delta^4(z-y), 
\end{equation}
which has the similar structure as eq.~(\ref{LG}). 
Assuming the translational invariance 
we set $ G(x,y) = G(x-y) $ and introduce the Fourier transform 
\begin{equation}
{\partial \over \partial x_0} G(x-y) = 
\int {d^4 p' \over (2\pi)^4} e^{ip'(x-y)} i p'_0 G(p'), 
\end{equation}
as eq.~(\ref{int-ip}). 
Performing the Fourier transform in eq.~(\ref{trans-inv-int}) 
we obtain 
\begin{equation}
\sum_{\mu=0}^3 k_\mu \Lambda_\mu^Q(p,p-k) = 
p_0 G(p) - (p_0 - k_0) G(p-k), 
\end{equation}
which becomes eq.~(\ref{WI-shift}) 
by shifting the four-momentum. 

Here we check the limiting case of eq.~(\ref{WI-Q}). 
If we replace the full Green function $G(p)$ 
by the free Green function $G_0(p)$ and set $k_0=0$, we obtain 
\begin{equation}
\sum_{\mu=1}^3 k_\mu \Gamma_\mu^Q(p+k,p) = 
- { p_0 \over m } {\vec k} \cdot ( {\vec p}+{{\vec k} \over 2} ), 
\end{equation}
and via $p_0 = - \epsilon$ this relation means 
\begin{equation}
\Gamma_\mu^Q(p,p) = \epsilon v_\mu, 
\end{equation}
where the right-hand side is the proper energy current vertex. 

The extension of this Ward identity to the case of Cooper pairs 
is also straightforward. 
The Ward identity for heat current vertex for Cooper pairs is 
\begin{equation}
\sum_{\mu=0}^3 k_\mu \Delta_\mu^Q(q+k,q) = 
q_0 D^{-1}(q+k) - (q_0+k_0) D^{-1}(q), \label{WIC-Q}
\end{equation}
where 
$\Delta_\mu^Q$ is the counterpart of $\Gamma_\mu^Q$. 

It should be noted that the factor $q_0$ represents 
the energy carried by a Cooper pair and 
is automatically taken into account by the commutation relation. 

\section{Conclusion}

I have explained the derivation of the Ward identity in detail. 
I hope that you can save time in understanding it by this APPENDIX. 

Our main results of Ward identities for Cooper pairs are 
eqs.~(\ref{WIC-e}) and (\ref{WIC-Q}) 
whose counterparts for electrons are eqs.~(\ref{WI-e}) and (\ref{WI-Q}). 
The charge and energy carried by Cooper pairs 
are properly taken into account by the commutation relations. 


\end{document}